# Global Inflation Dynamics: regularities & forecasts

***Akaev, Askar***, Prigogine Institute of Mathematical Investigations of Complex Systems, Moscow State University; "Complex System Analysis and Mathematical Modeling of the World Dynamics" Project, Russian Academy of Sciences; the First President of the Kyrgyz Republic

***Korotayev, Andrey V.***, "Complex System Analysis and Mathematical Modeling of the World Dynamics" Project, Russian Academy of Sciences; Institute of Oriental Studies and Institute for African Studies, Russian Academy of Sciences, Moscow; National Research University "Higher School of Economics"; Russian State University for the Humanities, Faculty of Global Studies of the Moscow State University

***Fomin, Alexey***, "Complex System Analysis and Mathematical Modeling of the World Dynamics" Project, Russian Academy of Sciences

---

***Abstract***

The analysis of dollar inflation performed by the authors through the approximation of empirical data for 1913–2012 with a power-law function with an accelerating log-periodic oscillation superimposed over it has made it possible to detect a quasi-singularity point around the $17^{th}$ of December, 2012. It is demonstrated that, if adequate measures are not taken, one may expect a surge of inflation around the end of this year that may also mark the start of stagflation as there are no sufficient grounds to expect the re-start of the dynamic growth of the world economy by that time. On the other hand, as the experience of the 1970s and the 1980s indicates, the stagflation consequences can only be eliminated with great difficulties and at a rather high cost, because the combination of low levels of economic growth and employment with high inflation leads to a sharp decline in consumption, aggravating the economic depression. In order to mitigate the inflationary consequences of the explosive growth of money (and, first of all, US dollar) supply it is necessary to take urgently the world monetary emission under control. This issue should become central at the forthcoming G8 and G20 summits.

Consider global inflation dynamics in 1980–2011 (see Fig. 1):

**Fig. 1.** World Consumer Price Inflation (% per year), 1980–2011

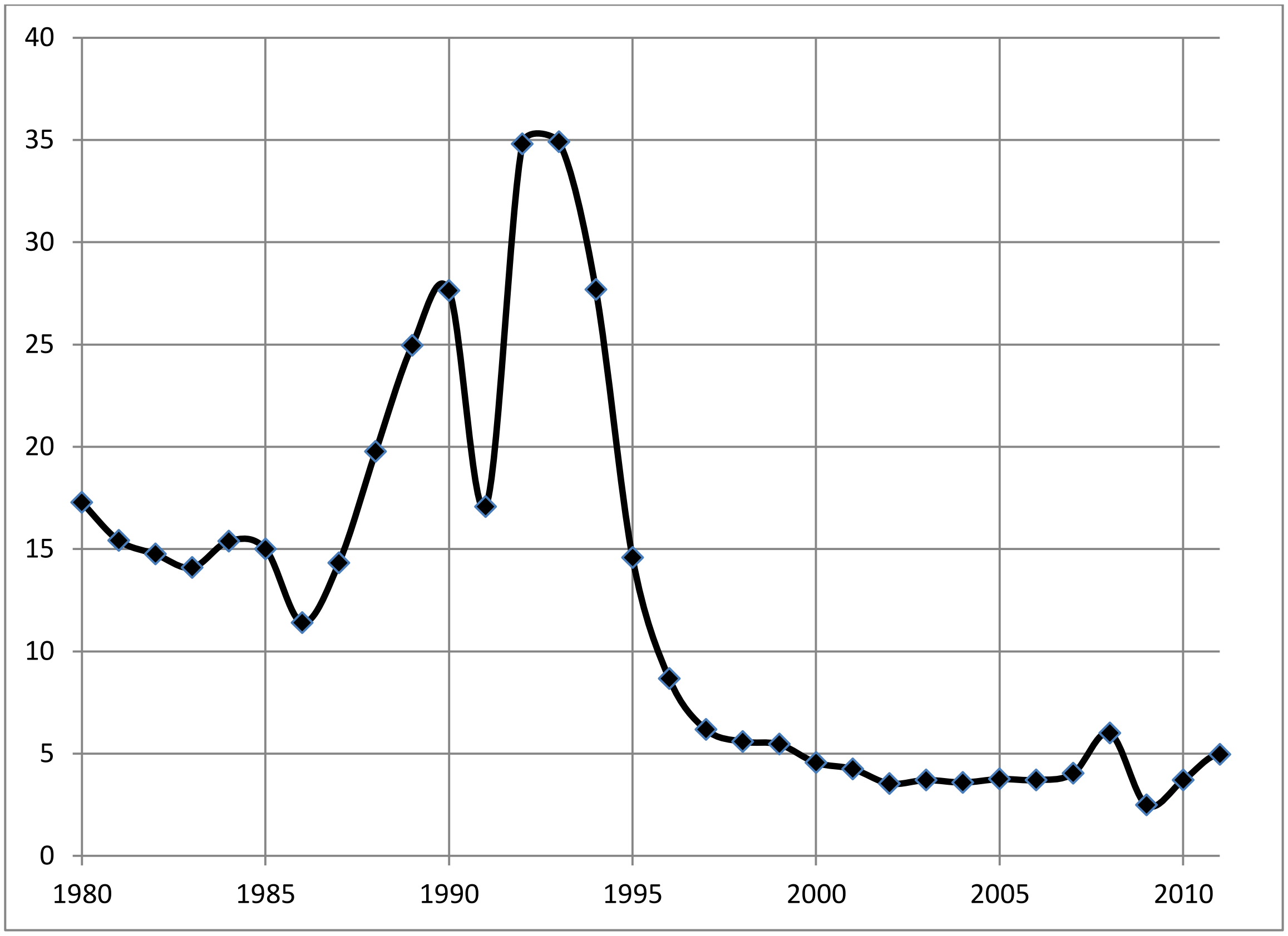

*Data source*: IMF 2012.

As can be seen, in 1980 global inflation was estimated to be around 17,3% per year; then, by 1986, it declined to 11%; yet, it started growing afterwards reaching it maximum value (35,3%) in 1992. Then it re-started its decline and by 2000 the world entered a phase of a relative consumer price stability. It should be noted that in the high-income OECD economies the price stabilization took place even earlier, by 1995. By the early 2000s the global inflation declined to the 4% level; it looked as if the World System had entered a golden era of price stability.

This era appears to be coming to its end. The point is that in the long-term perspective the global inflation depends primarily on the money supply by the USA, the producer of the main reserve currency in the world, the US dollar. The excessive supply of this currency observed in the recent years is capable of exploding the global price stability that still possesses certain inertia.

The low inflation rates of the recent decade turned out to have such strong inertia that it was not ended even by the explosive oil and raw commodity price growth of 2002–2008, whereas the “oil shocks” of the 1970s led to the bursts of very high inflation. Fig. 2 depicts inflation dynamics (calculated by the World Bank on the GDP deflator basis) for the world, for the high-income OECD countries, as well as for low- and middle-income economies:

**Fig. 2.** Inflation Rates in the World, in High-Income OECD Countries, as well as in Low- and Middle-Income Economies (% per year)

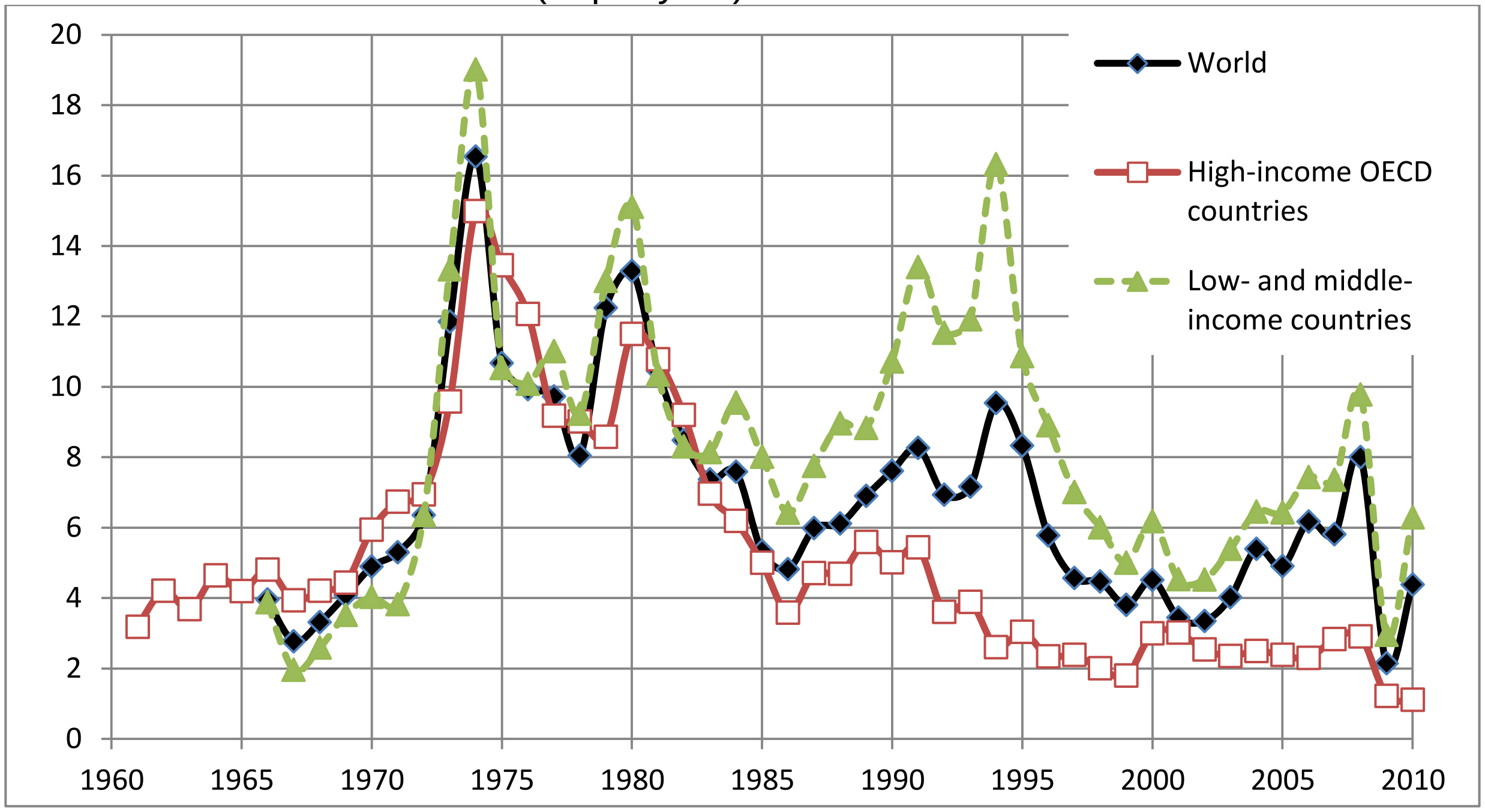

*Data source*: World Bank 2012: NY.GDP.DEFL.KD.ZG.

This diagram demonstrates that the oil price shocks of 1973 and 1979 led to rather strong bursts of inflation both in high- and low-/middle-income economies. However, in 2007–2008 inflation rates in the high-income countries hardly reacted to an oil-price shock quite comparable in its magnitude to the one of 1973, whereas the low- and middle-income economies reacted much less strongly than in the 1970s. This can be accounted for by a number of factors, including the growing use of alternative energy sources, energy-saving technologies, increase in the effectiveness of energy consumption; and one, of course, should not neglect the growth of the effectiveness of the monetary-credit policies.

On the other hand, during four decades after the collapse of the Bretton Wood International Monetary System the world was flooded with highly liquid assets that stimulated fast economic growth. One could also observe an explosive growth of international reserves, because with the disappearance of the Bretton Wood System its checks also disappeared, the mechanisms of self-regulation (that prevented automatically continuous distortions of trade balances) stopped working. This resulted in fast growth of trade and budget deficits.

The collapse of the Bretton Wood System made it possible for the USA to pay for imports with dollars not secured with anything (or with debt instruments valuated in dollars). One could observe the advent of the era of "paper money", and the volume of US dollars in circulation started growing explosively. Fig. 13 below demonstrates this in a rather salient way. Since 1970 the USA has accumulated more than $15 000 000 000 000 of its public debt. However, the explosive growth of the USA public debt cannot continue forever. Already today the USA is the largest debtor in the world history. Sooner or later the USA will lose its creditworthiness, which will mean the crisis of the US dollar. The fate of the dollar may be decided quite soon by the rate of the US budget deficit that reached an astronomic figure – $1 500 000 000 000 per year (see, *e.g.*, Executive Office of the President of the United States 2011); thus, one may expect a sharp decline of the dollar relative to other currencies and the gold. The price of gold may jump to $3000 per troy ounce. It is difficult to estimate how much the dollar will fall in relation to the other world currencies; however, this will impede substantially the continuation of the export-based world economic growth, whereas the USA trade deficit will not be able to continue working as a motor of the world economic growth (as this was observed in the recent three decades).

There are a number of factors that may unleash the spiral of inflation. First of all, this is the costs' inflation that has been caused in the recent years by the explosive growth of prices of production resources – oil, metals, other raw materials (see Fig. 3):

**Fig. 3.** IMF (Raw) Commodity Price Index (100 = 2005 level)

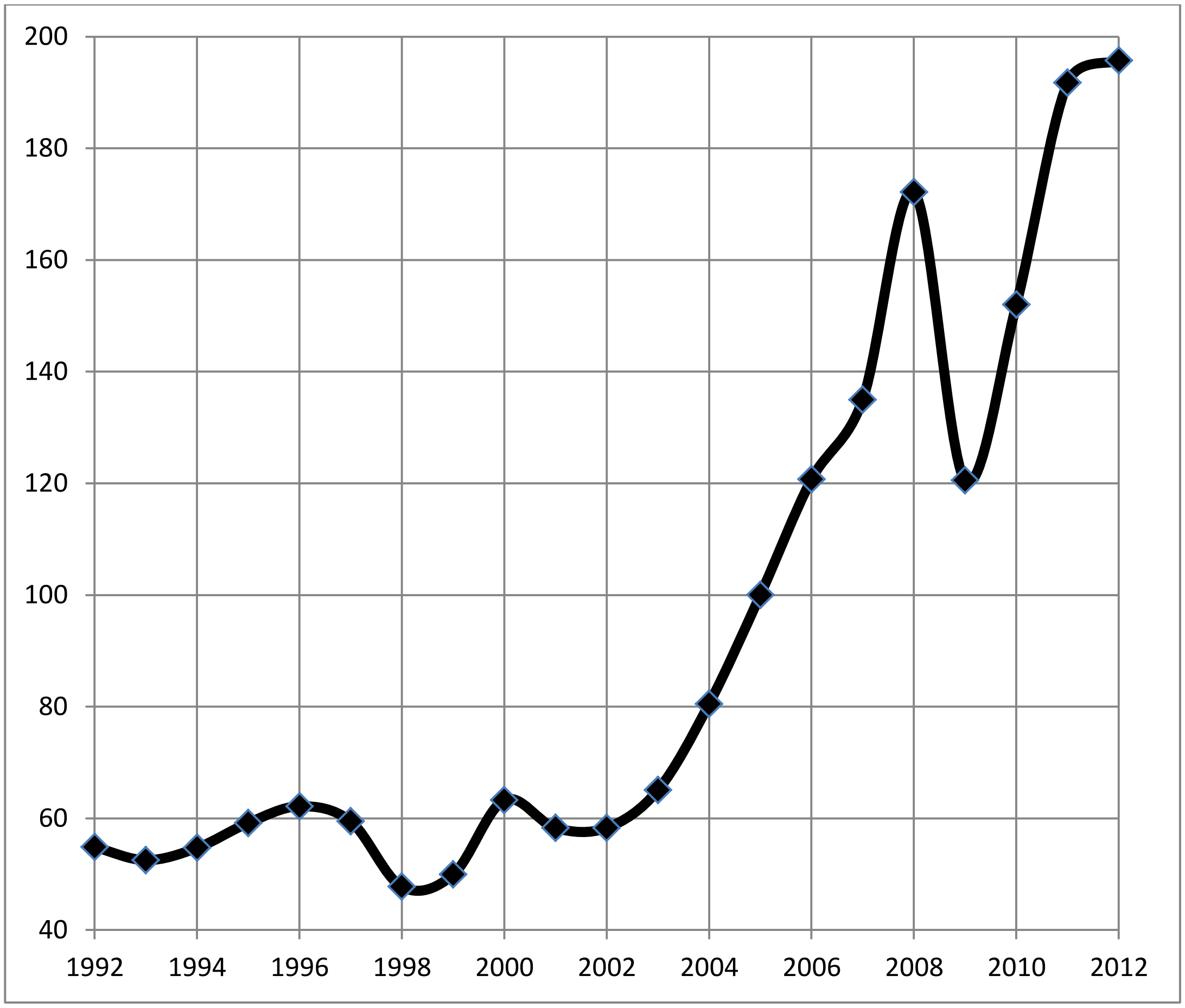

*Note:* Commodity Price Index is calculated by the IMF taking into account dynamics of prices of fuels, industrial metals and the main agricultural resources (wheat, corn, rice, sugar, cooking oils, etc.). *Data source*: IMF 2012. 2011 – IMF estimate; 2012 – IMF forecast.

Consider now how in 2002–2008 the main economic actor of the modern World System (the USA) managed to avoid any significant growth of its consumer price index notwithstanding so dramatic increase in the costs' inflation that was observed in those years. In order to do this, consider simultaneous dynamics of three indicators in those years – the world Commodity Price Index, the USA Producer Price Index, and the USA Consumer Price Index (Fig. 4):

**Fig. 4.** Dynamics of the World (Raw) Commodity Price Index, US Producer Price Index and US Consumer Price Index (100 = 2002 level)

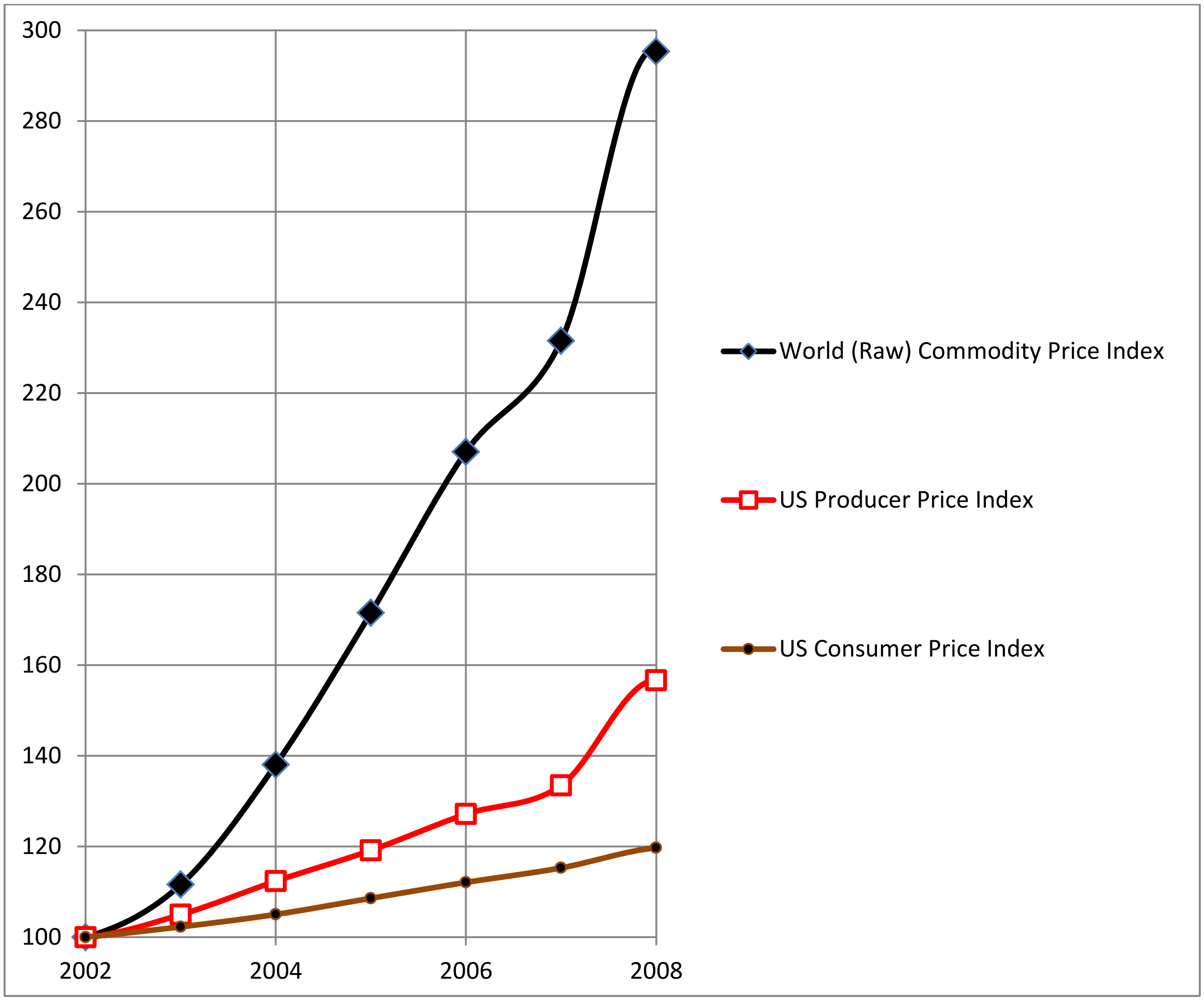

*Data sources*: IMF 2012; World Bank 2012: FP.CPI.TOTL; Bureau of Labor Statistics 2012.

As we see, in 2002–2008 (that is, during the period of the fastest cost inflation) the American producers managed to achieve impressive successes in the containment of the price growth. Indeed, during those years the prices of fuels and basic raw materials grew almost three times; in the meantime the American Producer Price Index only grew by about 50%, and, of course, the growth of the effectiveness of the energy and raw material consumption by the American producers played here a rather important role. Yet, in the same period of time the US consumer price index only increased by less than 20%!

What could account for this discrepancy? Of course, the main point here is that during those very years the Americans bought more and more cheap goods produced in the World System periphery, in general, and in China, in particular. As a result, the American consumer price index grew much slower than the American producer price index – as every year the goods purchased by the American consumers became less and less American. Naturally, this process was directly connected with the explosive growth of the US import, in general, and the one from China, in particular (see Figs. 5–6):

**Fig. 5.** Overall US Import Dynamics (US dollars)

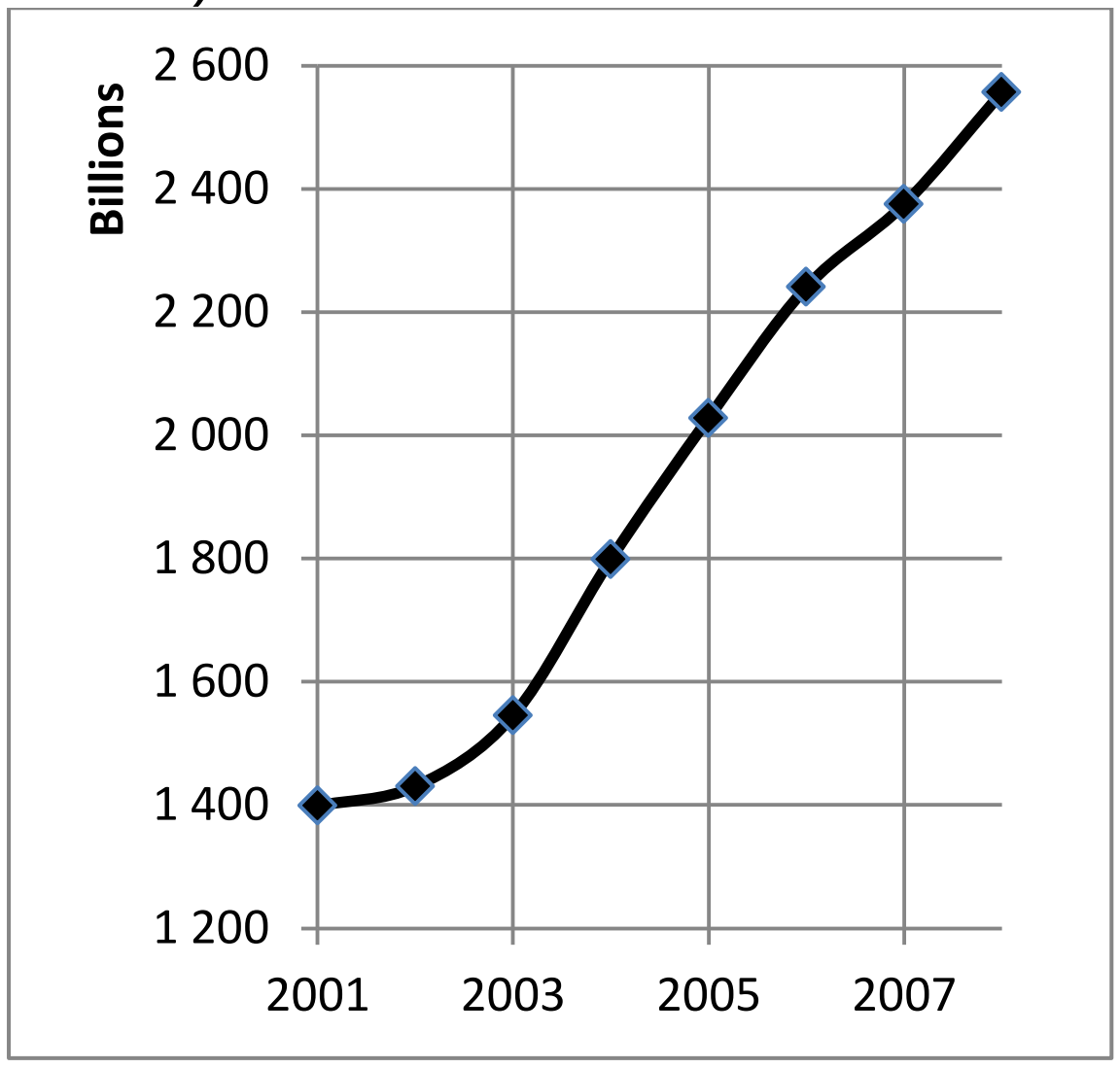

*Data source*: World Bank 2012: NE.IMP.GNFS.CD.

**Fig. 6.** Dynamics of the US Import from the People's Republic of China (US dollars)

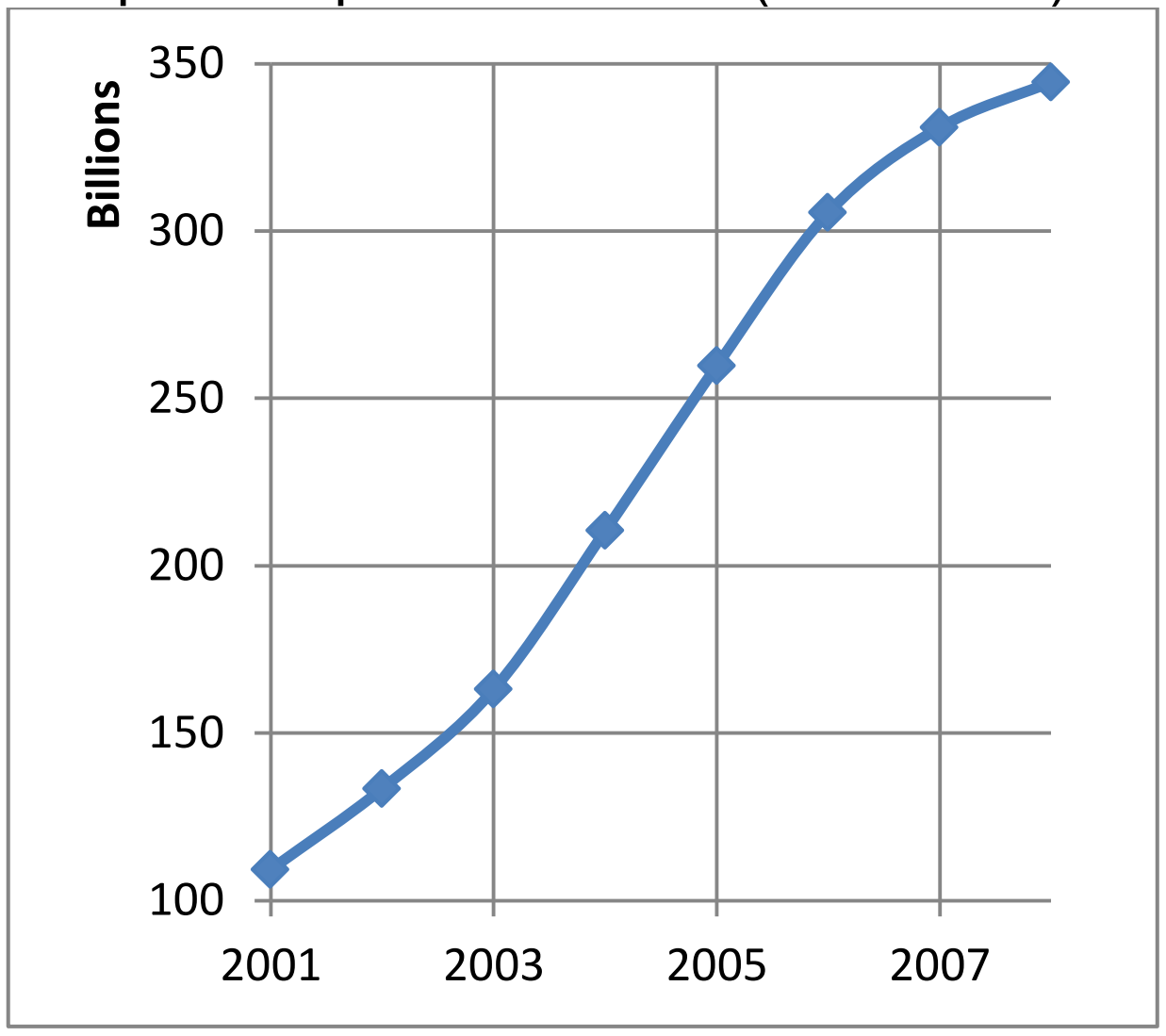

*Data source*: MIT 2012.

As we see, in 2001–2008 (that is, just within 7 years) the US import grew almost twice, whereas the US import from China grew more than three times.

It appears necessary to emphasize that the growth of the American export lagged far behind the explosive increase in the value of the US import, which, naturally resulted in a very fast growth of the US foreign trade deficit (but, in particular, the one of its trade with China) (see Figs. 7–8):

**Fig. 7.** Overall Dynamics of the US Foreign Trade Deficit (US dollars)

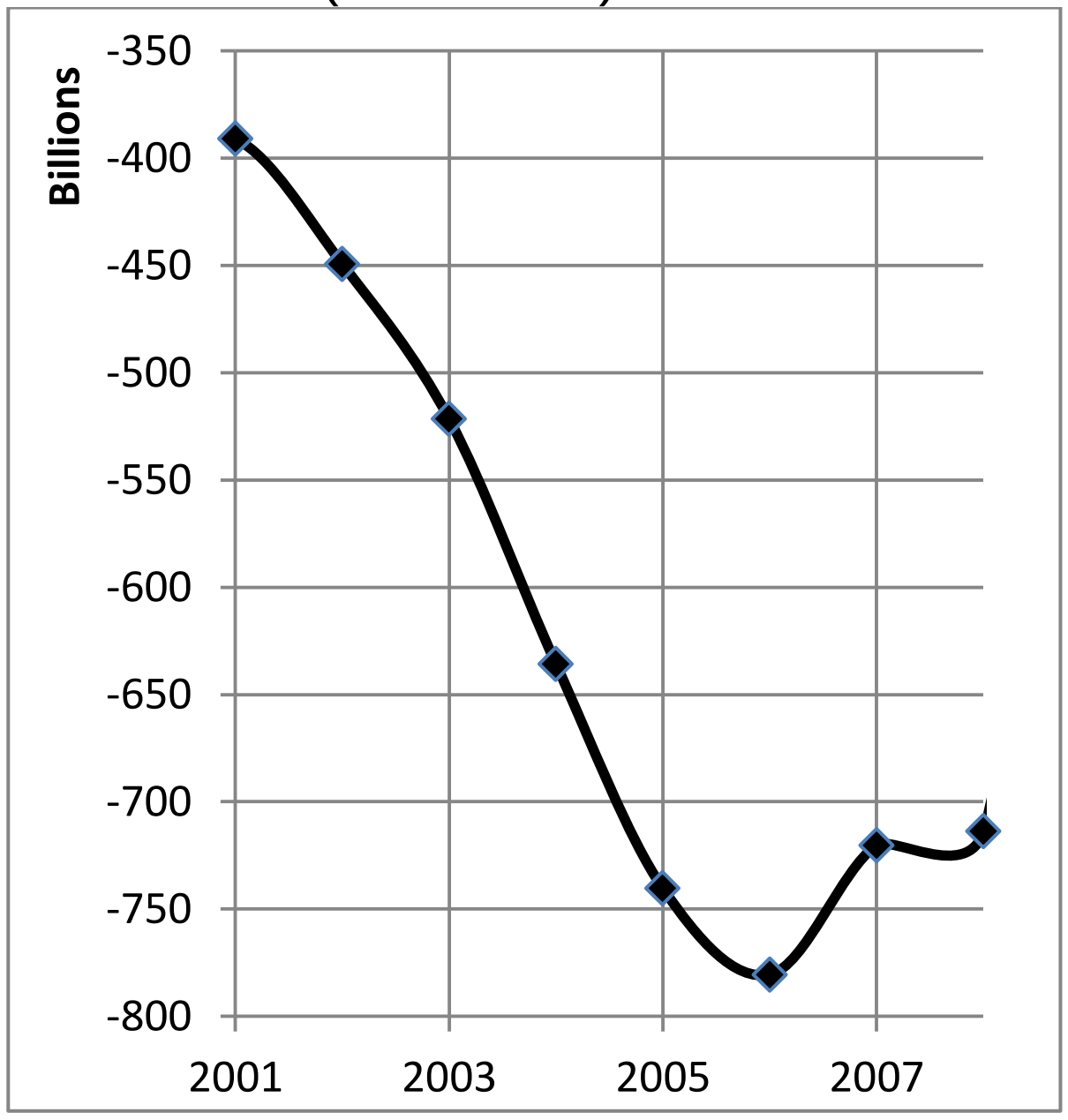

*Data source*: World Bank 2012: NE.IMP.GNFS.CD; BX.GSR.GNFS.CD.

**Fig. 8.** Dynamics of the Deficit of the US Trade with China (US dollars)

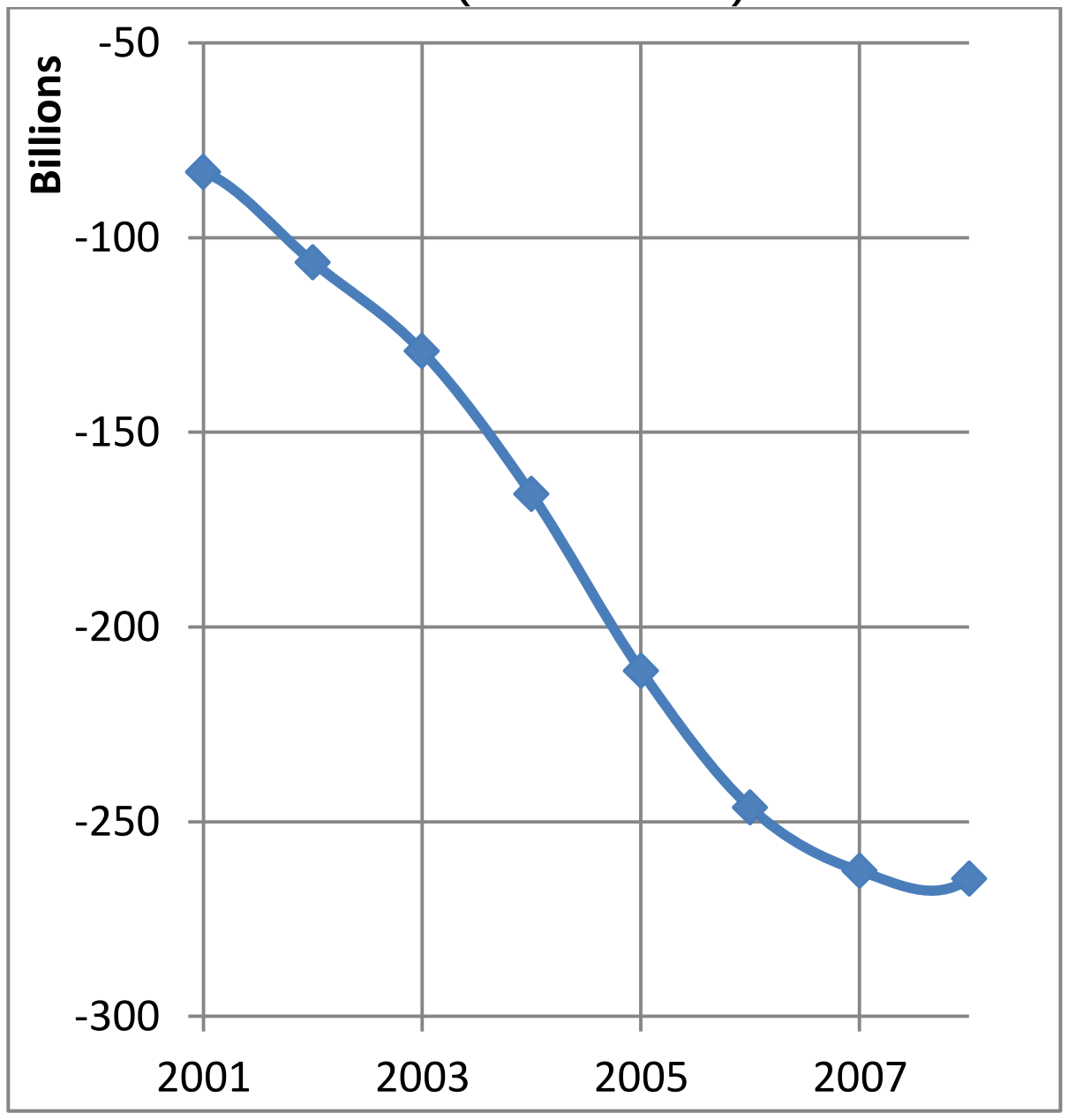

*Data source*: MIT 2012.

As we see, we can observe here a situation that is strikingly similar to the one that we have already seen as regards the US import – in 2001–2008 the US foreign trade deficit grew almost twice, whereas this deficit in trade with China increased more than three times. As a result, in 2008 almost 40% of the US foreign trade deficit was accounted for by its trade with the People's Republic of China (against this background, it does not appear coincidental that at present China is the largest holder of the US public debt [U.S. Department of the Treasury 2012]).

It does not seem to be coincidental either that in the same period of time we observe the collapse of US Federal Budget deficit figures to 11-digit levels (see Fig. 9):

**Fig. 9.** USA Federal Budget Deficit Dynamics (dollars)

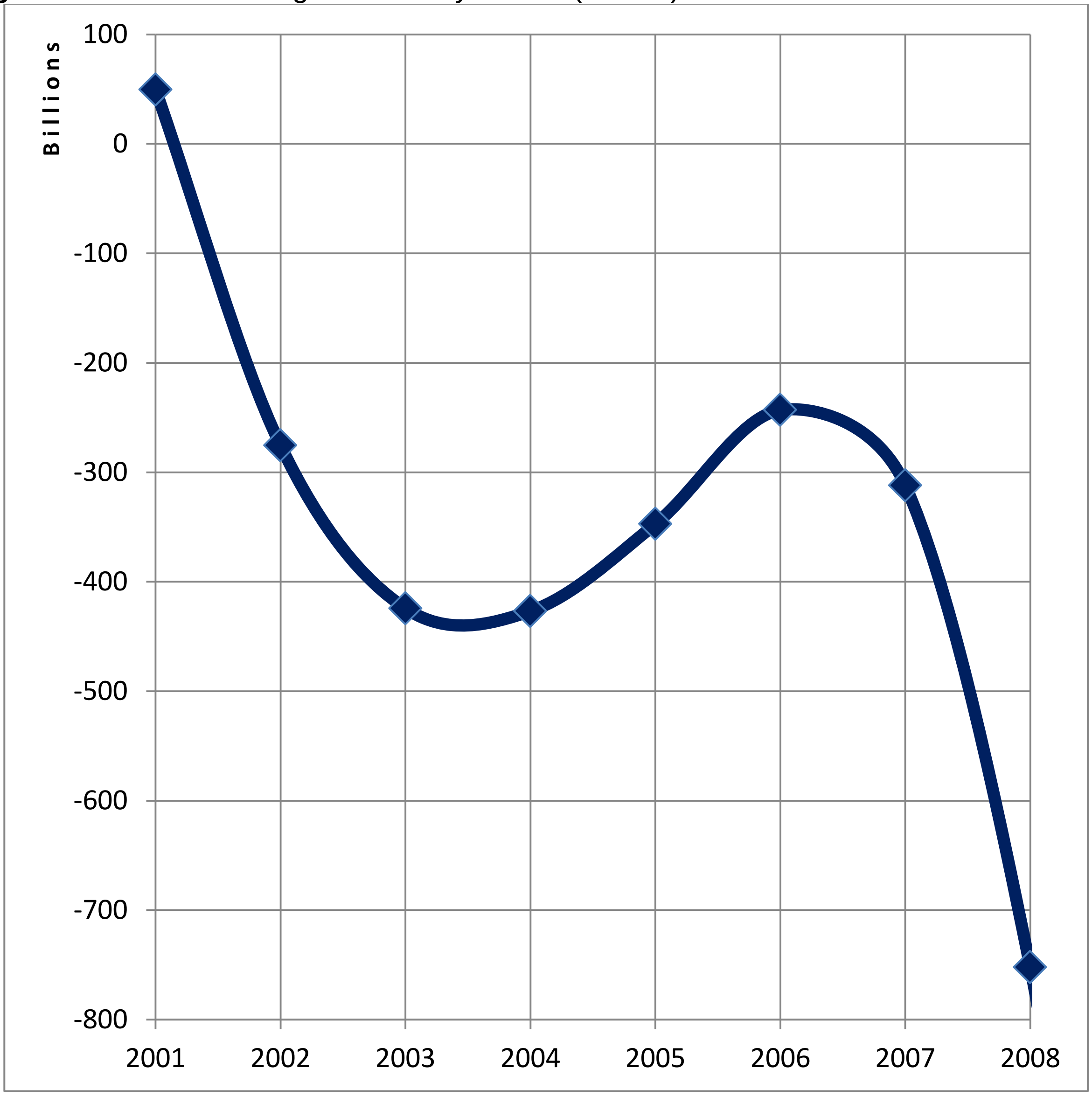

*Data source*: World Bank 2012: GC.BAL.CASH.CN.

In the same years one could also observe an explosive growth of the US public debt (see Fig. 10):

**Fig. 10.** US Public Debt Dynamics (millions of US dollars)

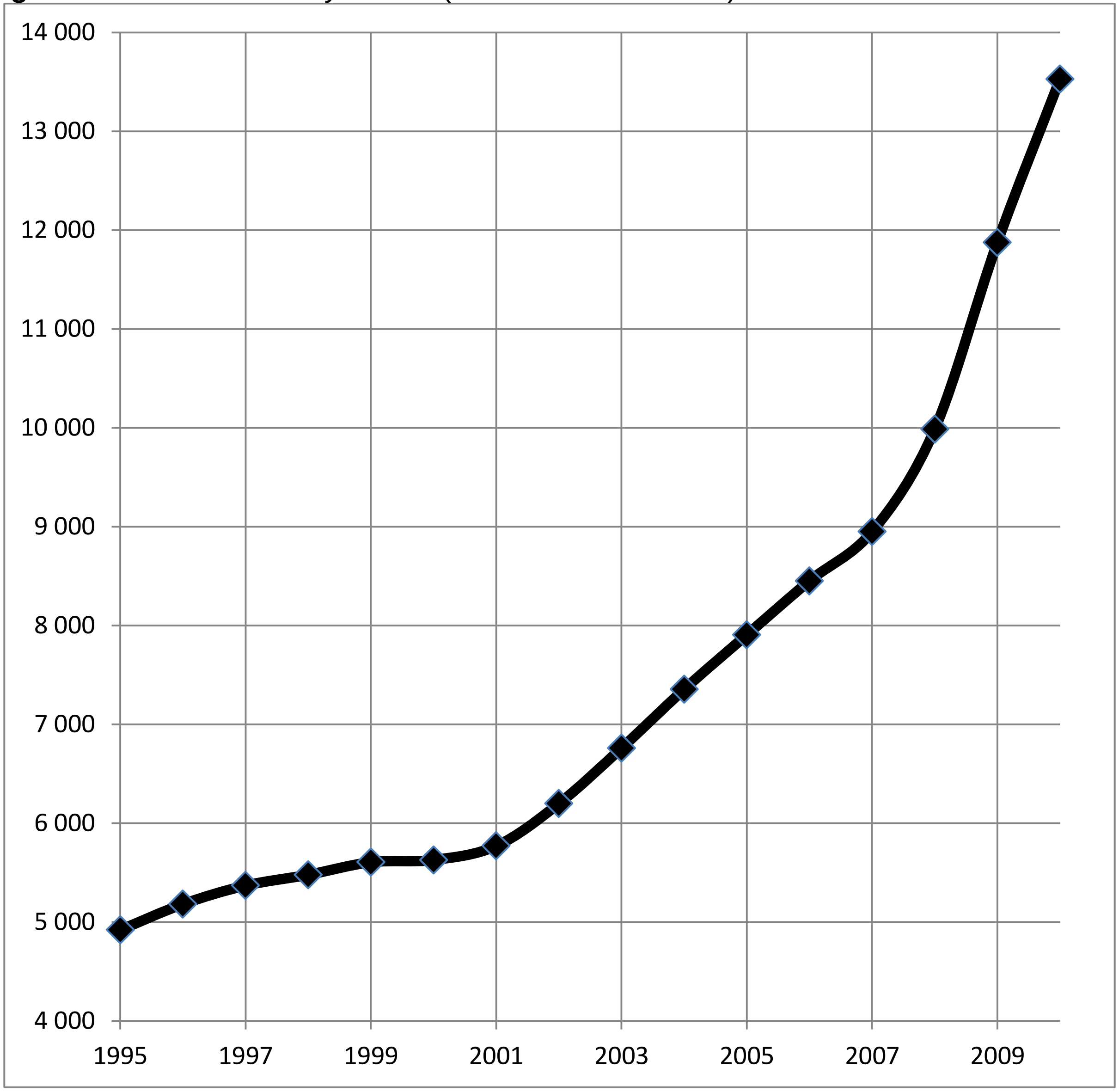

*Data source*: Federal Reserve Bank of St. Louis 2012.

Thus, in order to preserve low consumer inflation level the USA paid a dramatic cost – it managed to do this at the expense of getting in a sort of debtor's hole (let alone the country deindustrialization – decline of industrial production, or even close-down of those American industrial enterprises that were unable to compete with the cheap Chinese imports).

It does not appear strange at all that those processes (as we will see below) were accompanied in the USA by an explosive growth of the M2 aggregate mass (which, after the start in 2008 of the acute phase of the world crisis was supplemented by a very fast increase in the M1 aggregate mass).

In any case we observe a powerful monetary inflation factor connected with the fast emission of the world currencies – first of all the US dollar, but also yuan, euro and so on. The tremendous increase in money supply is one of the main sources of inflation for both developed and developing countries (see Fig. 11):

**Fig. 11.** Dynamics of Overall Monetary Mass M1, 2003 – February 2012, 100 = 2005 level

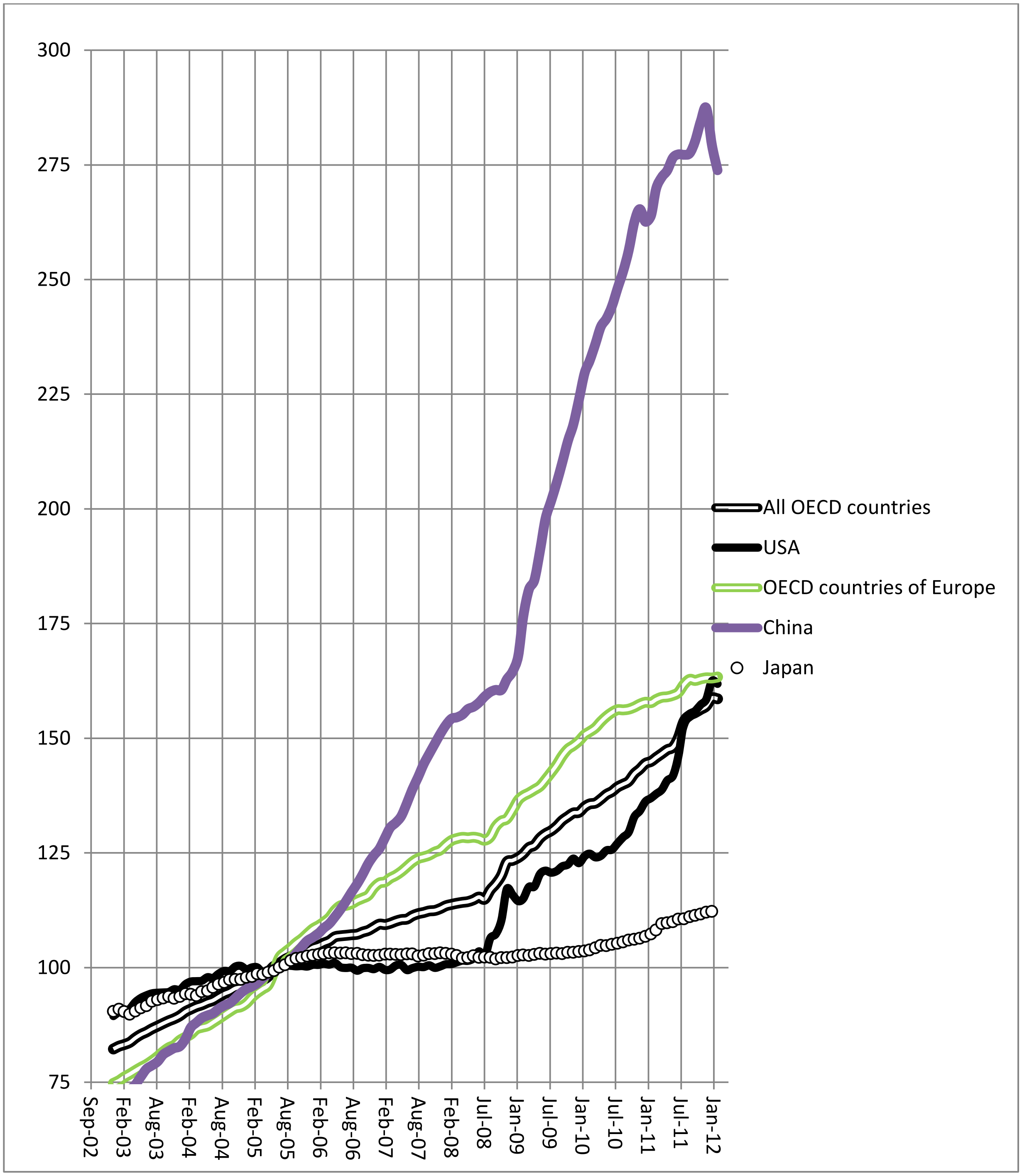

*Source*: OECD 2012.

However, among all the macroeconomic actors a really explosive growth of the M1 monetary mass is only observed in the USA (see. Fig. 12):

**Fig. 12.** M1 Monetary Mass Dynamics in the USA, 2005 – March 2012, billions of dollars

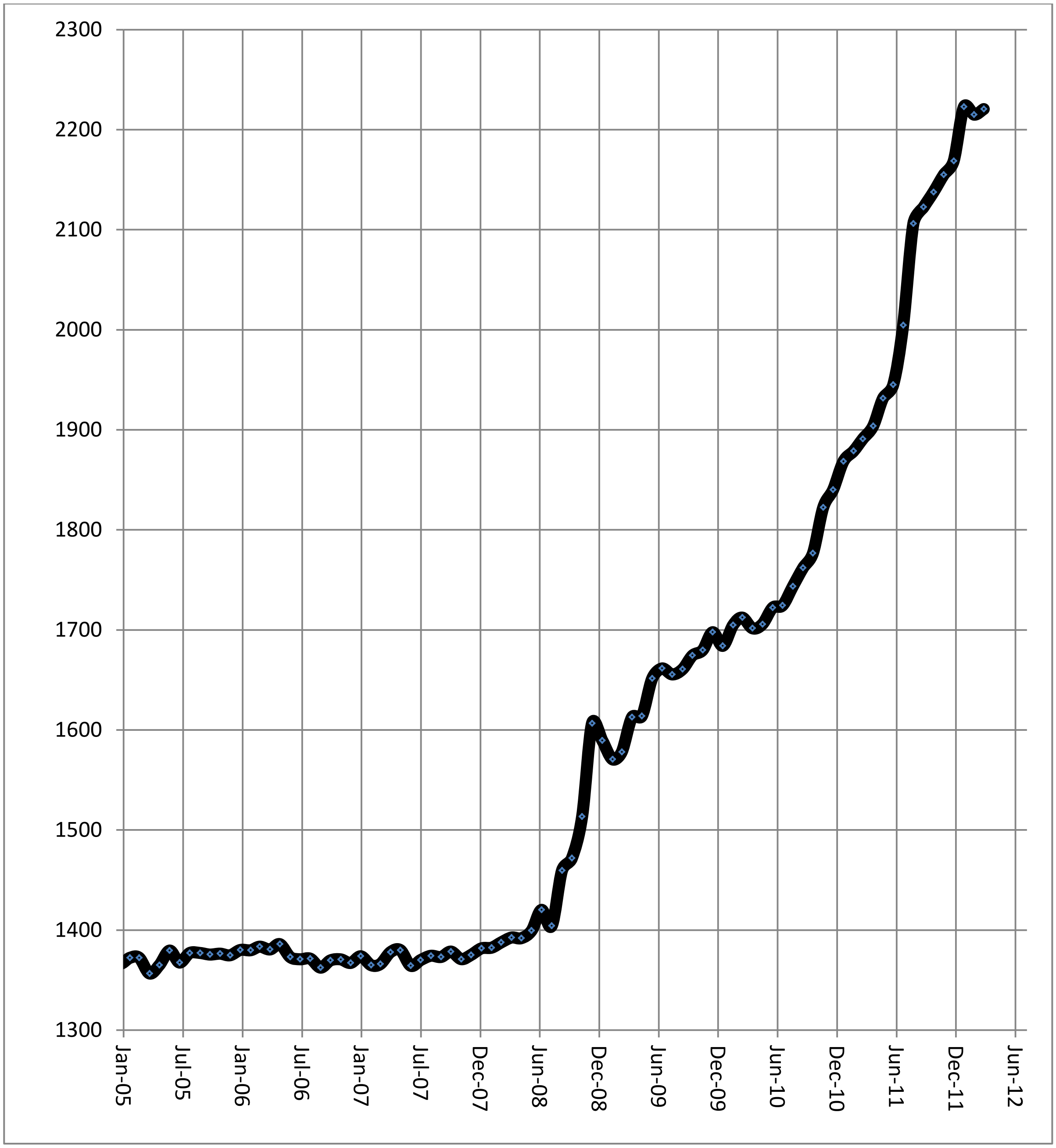

*Data source*: Board of Governors of the Federal Reserve System 2012.

However, even an extremely fast growth of the American M1 monetary mass looks like a pale reflection of a really explosive growth of the US M2 aggregate (Fig. 14).

Note that according to the Friedman – Phelps Monetary Model (Самуэльсон, Нордхаус 2009: 460), in short term the pumping of liquidity helps to increase economic growth rates and to decrease unemployment; however, in the long term it generates inflation. It appears that at present we are rather close to the respective moment.

The excessive money supply is directly connected with the Federal Budget deficit, as recently this deficit was covered to a considerable extent through the emission of money and governmental debt instruments. Sargent and Wallace (1981) come to the conclusion (while modeling inflation processes) that the general growth of prices is generated by the emission of both money and debt instruments. What is more, they demonstrate that, in the long term the emission of the government debt instruments may have more substantial inflationary consequences. Thus, in long term, the inflation levels are determined first of all by the state of the governmental budget.

**Fig. 13.** M1 and M2 Monetary Mass Dynamics in the USA, 1959 – March, 2012, billions of dollars

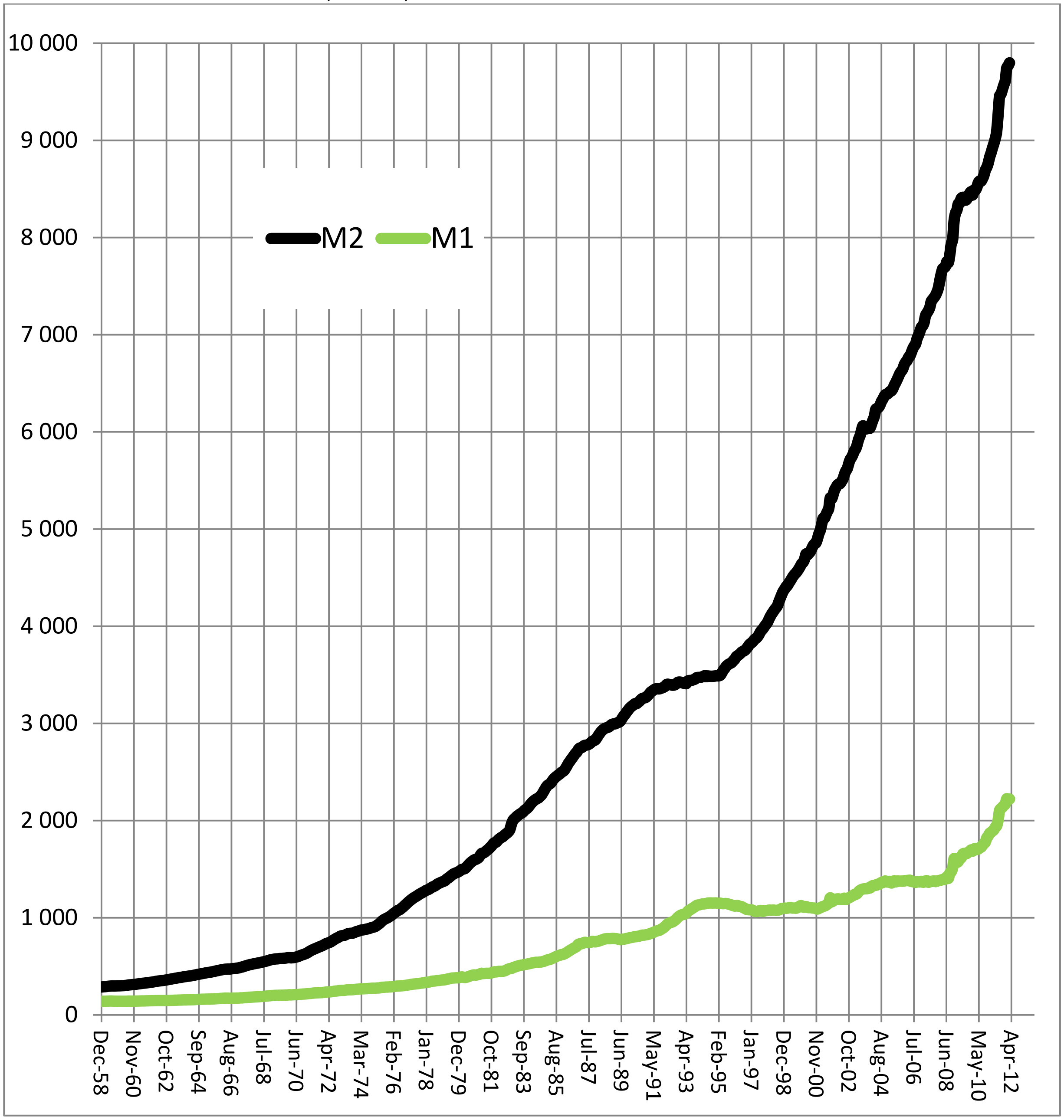

*Data source*: Board of Governors of the Federal Reserve System 2012.

Thus, on the basis of Friedman's model, Bruno and Easterly formulate the following thesis: "High inflation always and everywhere is connected with the budget deficit" (Моисеев 2004: 60). As is well known, all the developed countries (including the USA) currently suffer from budget problems, high deficits and public debts (see, *e.g.*, IMF 2012) whose servicing becomes an intolerable burden for real economy. The monetary emission has become today a widely used method to finance the governmental spending, whereas previously in the developed countries (where rather effective taxation systems and financial markets used to be functioning) seigniorage was hardly used in any extensive way for fiscal purposes.

In the present-day world rates of money supply are growing rather fast, whereas in the USA we seem to be dealing with a sort of blow-up regime.

* * *

In a number of seminal works by Didier Sornette, Anders Johansen and their colleagues (Sornette, Sammis 1995; Sornette, Johansen 1997, 2001; Johansen, Sornette 1999, 2001; Johansen *et al.* 1996; Sornette 2004; etc.) it has been demonstrated that accelerating log-periodic oscillations superimposed over an explosive growth trend that is described with a power-law function with a singularity (or quasi-singularity) in a finite moment of time $t_c$,

$$x(t) = A - m\,(t_c - t)^{\alpha}\,\{\,1 + C\cos[\omega\ln(t_c - t) + \varphi]\,\}, \tag{1}$$

are observed in situations leading to crashes and catastrophes. They can be analyzed because their precursors allow the forecasting of such events. One can mention such examples as the log-periodic oscillations of the Dow Jones Industrial Average (DJIA) that preceded the crash of 1929 (*e.g.*, Sornette, Johansen 1997), or the changes in the ion concentrations in the underground waters that preceded the catastrophic Kobe earthquake in Japan on the 17th of January, 1995 (*e.g.*, Johansen *et al.* 1996), which are also described mathematically rather well with log-periodic fluctuations superimposed over a power-law growth trend.

Equation (1) describes power-law growth with log-periodic oscillations (described by the term $C\cos[\omega\ln(t_c - t) + \varphi]$) superimposed over it. Log-oscillations are designated this way because in the logarithmic scale – when we use $\ln(t_c - t)$ values as marks at the time axis – those oscillations look like periodic. An important feature of those oscillations is their fractal character – one could observe the superimposition over the parametrization with the abovementioned equation of smaller log-oscillations that are still described with the same equation – $C\cos[\omega\ln(t_c - t) + \varphi]$ ) – but with different parameters.

Our analysis of the American Producer Price Index (Bureau of Labor Statistics 2012) for the period between January 1913 and March 2012 with equation (1) has produced the following results (see Fig. 14):

**Fig. 14.** American Producer Price Index Dynamics
(January, 1913 – March, 2012; 100 = 1982 level)

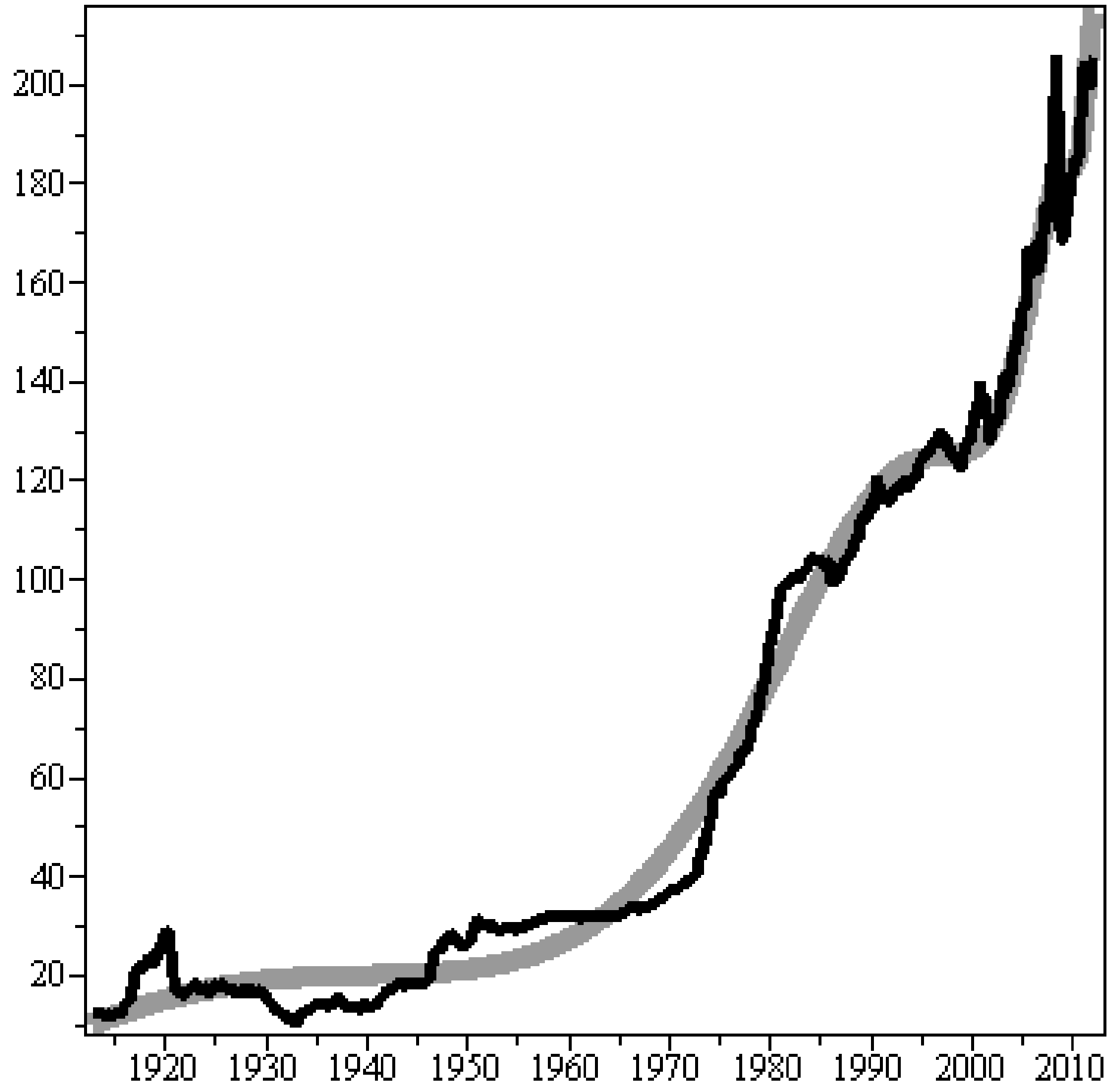

*Note*: the black curve corresponds to empirical estimates of the US Bureau of Labor Statistics (2012); the grey curve was generated by equation (1) with the following parameter values: $A$ = 252.7; $m$ = 44.01; $C$ = 0.092; $t_c$ = 2012.965; $\alpha$ = 0.392; $\omega$ =9.46; $\varphi$ = 2.96. All the parameters (excluding the singularity point) were calculated through the dispersion minimization method. The quasi-singularity point was calculated with the successive exclusion method (see Акаев, Коротаев, Фомин 2012: Appendix) as 2012.965, which corresponds to the 17th of December 2012.

Thus, our calculations have indicated the 17th of December, 2012 as the singularity point. This implies that the Federal Reserve System must change its monetary-credit policy in a rather radical way. The FRS cannot continue an explosive emission of unsecured dollar mass in order to finance the enormous governmental budget deficit of the USA (around 10% of its GDP [see, *e.g.*, Executive Office of the President of the United States 2011: 171]) and to devaluate the overgrown US public debt that has already exceeded substantially 100% of its GDP (see, *e.g.*, IMF 2012). The USA will have to achieve the restoration of the equilibrium of its international balance of payment (and this is most likely to occur as a result of the US dollar collapse).

The forecasted dollar crisis is likely to act as that very shock that will explode the price stability at the global level and will unleash the spiral of global inflation. Thus, if adequate measures are not taken, one may expect a surge of inflation around the end of this year that may also mark the start of stagflation as there are no sufficient grounds to expect the re-start of the dynamic growth of the world economy by that time (see, e.g., Akaev, Sadovnichy, Korotayev 2012; Akaev, Fomin, Korotayev 2011; Korotayev, Tsirel 2010; Korotayev, Zinkina, Bogevolnov 2011). On the other hand, as the experience of the 1970s and the 1980s indicates, the stagflation consequences can only be eliminated with great difficulties and at a rather high cost, because the combination of low levels of economic growth and employment with high inflation leads to a sharp decline in consumption, aggravating the economic depression.

Hence, in order to mitigate the inflationary consequences of the explosive growth of money (and, first of all, US dollar) supply it is necessary to take urgently the world monetary emission under control. This issue should become central at the forthcoming G8 and G20 summits.

## Bibliography

**Akaev A., Fomin A., Korotayev A. 2011.** The Second Wave of the Global Crisis? On mathematical analyses of some dynamic series. *Structure & Dynamics* 5/1: 1–10.

**Akaev A., Sadovnichy A., Korotayev A. 2012.** On the dynamics of the world demographic transition and financial-economic crises forecasts. *The European Physical Journal* 205: 355-373.

**Board of Governors of the Federal Reserve System. 2012.** *Money Stock Measures.* URL: http://www.federalreserve.gov/releases/H6/default.htm.

**Bureau of Labor Statistics. 2012.** *Producer Price Indexes. PPI Databases.* Washington, DC: United States Department of Labor. URL: http://www.bls.gov/ppi/data.htm.

**Executive Office of the President of the United States. 2011.** *Fiscal Year 2012 Budget of the U.S. Government*. Washington, DC: U.S. Government Printing Office.

**Federal Reserve Bank of St. Louis. 2012.** *FRED Economic Data. Gross Federal Debt*. URL: http://research.stlouisfed.org/fred2/series/FYGFD/downloaddata?cid=5.

**IMF (International Monetary Fund). 2012**. *World Economic Outlook. Growth Resuming, Dangers Remain. World Economic Outlook Database.* Washington, DC: International Monetary Fund. URL: http://www.imf.org/external/pubs/ft/ weo/2012/01/weodata/index.aspx.

**Johansen A., Sornette D. 1999.** Critical Crashes. *Risk* 12/1: 91–94.

**Johansen A., Sornette D. 2001.** Finite-time Singularity in the Dynamics of the World Population and Economic Indices. *Physica A* 294/3–4: 465–502.

**Johansen A., Sornette D., Ledoit O. 1999.** Predicting financial crashes using discrete scale invariance. *Journal of Risk* 1/4: 5–32.

**Johansen A., Sornette D., Wakita H., Tsunogai U., Newman W. I., Saleur H. 1996.** Discrete scaling in earthquake pre-cursory phenomena: Evidence in the Kobe earthquake, Japan. *Journal de Physique I* 6/10: 1391–1402.

**Korotayev A., Tsirel S. 2010.** A Spectral Analysis of World GDP Dynamics: Kondratieff Waves, Kuznets Swings, Juglar and Kitchin Cycles in Global Economic Development, and the 2008–2009 Economic Crisis. *Structure and Dynamics* 4/1: 3–57. URL: http://www. escholarship.org/uc/item/9jv108xp.

**Korotayev A., Zinkina J., Bogevolnov J. 2011.** Kondratieff Waves in Global Invention Activity (1900–2008). *Technological Forecasting & Social Change* 78: 1280–1284.

**MIT (Massachusetts Institute of Technology). 2012.** *The Observatory of Economic Complexity*. URL: http://atlas.media.mit.edu/explore/ tree_map/ import/ usa/chn/.

**OECD (= Organization for Economic Co-operation and Development). 2012.** *Monthly Monetary and Financial Statistics.* URL: http://stats.oecd.org/index.aspx?querytype=view&queryname=170#.

**Sargent T. J., Wallace N. 1981.** Some Unpleasant Monetarist Arithmetic. *Federal Reserve Bank of Minneapolis Quarterly Review* 5/3: 1–17.

**Sornette D. 2004.** *Why stock markets crash: critical events in complex financial systems.* Princeton, NJ: Princeton University Press.

**Sornette D., Johansen A. 1997.** Large financial crashes. *Physica A* 245/3–4: 411–422.

**Sornette D., Johansen A. 1998.** A hierarchical model of financial crashes. *Physica A* 261/3–4: 351–358.

**Sornette D., Johansen A. 2001.** Significance of log-periodic precursors to financial crashes. *Quantitative Finance* 1/4: 452–471.

**Sornette D., Sammis C. G. 1995.** Complex critical exponents from renormalization group theory of earthquakes: Implications for earthquake predictions. *Journal de Physique I* 5/5: 607–619.

**Sornette D., Woodard R., Zhou W.-X. 2009.** The 2006–2008 Oil Bubble: evidence of speculation, and prediction. *Physica A* 388: 1571–1576.

**U.S. Department of the Treasury. 2012.** *Major Foreign Holders of Treasury Securities*. URL: http://www.treasury.gov/resource-center/data-chart-center/tic/Documents/mfh.txt

**World Bank. 2012.** *World Development Indicators Online.* Washington, DC: World Bank. URL: http://data.worldbank.org/indicator.

**Акаев А. А., Коротаев А. В., Фомин А. А. 2012.** *Динамика темпов глобальной инфляции: закономерности и прогнозы.* М.: УРСС.

**Моисеев С. Р. 2004.** *Инфляция: современный взгляд на вечную проблему*. М: Маркет ДС.

**Самуэльсон П. Э., Нордхаус В. Д. 2009.** *Макроэкономика*. М: ООО «И. Д. Вильямс».